\documentclass[draftcls,onecolumn]{IEEEtran}

\usepackage[utf8]{inputenc}
\usepackage{mathtools,amssymb}
\usepackage{graphicx}
\usepackage{mathrsfs}

\DeclareRobustCommand*{\IEEEauthorrefmark}[1]{%
  \raisebox{0pt}[0pt][0pt]{\textsuperscript{\footnotesize #1}}%
}

\begin{document}
\title{Two-step Machine Learning Approach for Channel Estimation with Mixed Resolution RF Chains}

\author{
\IEEEauthorblockN{\textit{Brenda Vilas Boas\IEEEauthorrefmark{1}$^,$\IEEEauthorrefmark{2}, Wolfgang Zirwas\IEEEauthorrefmark{1}},
\textit{Martin Haardt\IEEEauthorrefmark{2}}}\\\ 
\IEEEauthorblockA{\IEEEauthorrefmark{1}Nokia Bell Labs Munich, Germany \\\ \IEEEauthorrefmark{2}Ilmenau University of Technology, Germany }}

\maketitle
\begin{abstract}
Massive MIMO is one of the main features of 5G mobile radio systems. However, it often leads to
high cost, size and power consumption. To overcome these issues, 
the use of constrained radio frequency (RF) frontends has been proposed, as well as novel precoders, 
e.g., a  
multi-antenna, greedy, iterative and quantized precoding algorithm (MAGIQ). Nevertheless, the best performance of MAGIQ
assumes accurate channel knowledge per antenna element, for example, from uplink sounding reference signals. 
In this context, we propose an efficient uplink channel estimator by applying machine learning (ML) algorithms. 
In a first step a conditional generative adversarial network (cGAN) predicts the radio channels 
from a limited set of full resolution RF chains to  
the rest of the low resolution RF chain 
antenna elements.
A long-short term memory (LSTM) neural network extracts further phase information from the low resolution RF chain antenna elements. Our results indicate that our proposed approach is competitive with traditional Unitary tensor-ESPRIT in scenarios with various closely spaced multipath components (MPCs).

\end{abstract}


\section{Introduction}

Consider a massive 
MIMO system, where every antenna element is connected to a dedicated radio frequency (RF) chain including a full resolution analog-to-digital converter (ADC). 
With increasing number of antenna elements, the energy consumption becomes a serious implementation issue. 
In this regard, the usage of constrained RF frontends has been proposed in~\cite{17ZirwascMIMO} and later enhanced by suitably adapted downlink multi-user (MU)
MIMO precoders, such as
the MAGIQ algorithm~\cite{18ZirwasQuantized}. MAGIQ relies on accurate channel state information (CSI), which generally means one full receiver RF chain per antenna element. Receiver RF chains are less complex than transmiter chains; however, limiting the uplink receiver complexity is also important to reduce cost and size.
Then, a mix of full and low resolution ADCs can provide a reasonable trade-off between performance and power consumption. Here, we analyze suitable machine learning (ML) methods to  most accurately infer the CSI for the low
to the high resolution RF
chains.    

Generative adversarial networks (GANs) were first proposed in ~\cite{14GoodfellowGAN}, where 
a min-max game is played between two neural networks (NNs), named generator and discriminator networks, aiming to train the generator to output realistic images from random noise at its input. Since then, many enhancements have been proposed to the pioneering GAN architecture with most of its success on image reconstruction, image super-resolution, and image domain translation~\cite{17IsolaPix,17LedigPhoto}. 

For wireless mobile radio systems, GANs are often concerned with physical layer issues like channel modeling 
and data augmentation~\cite{19YangGANmodeling,19OsheaVenc}. 
A conditional GAN (cGAN) is used in~\cite{18LiGANcov} to estimate the millimeter wave (mm-Wave) \textit{virtual} covariance channel matrix based on prior knowledge of a training sequence.          
A cGAN and a variational autoencoder (VAE) GAN are used in~\cite{18YeChannelE2E,19SmithCommDensity}, but in a context of end-to-end learning where the final objective is to predict the transmitted symbols, not the wireless channel.  
The authors in ~\cite{19TakedaNonIdealADCsDNNvsGAMP,20ZhangDNN1bitfewPilots} propose fully connected NNs for the problem of channel estimation for constrained  
massive MIMO systems with 1-bit ADCs. Their results are compared with variations of the generalized message passing (GAMP) algorithm. Moreover,~\cite{19GaoDNNmixedresADCs} proposes two fully connected NNs for channel estimation in a mixed scenario with full and low resolution ADCs. However, it concludes that the best strategy is one without taking into account 
the quantized signals from the low resolution ADCs due to their distortion and low resolution.

Motivated by the good results of GANs in image to image translation 
and by the challenge of wireless channel estimation with low resolution ADCs, in this paper we propose a 2-step ML
algorithm to be carried out by the base station (BS).  
 We first perform channel estimation considering just the full resolution measurements by using the \textit{Pix2Pix} GAN architecture ~\cite{17IsolaPix}, which is a type of cGAN. Second, we enhance the channel estimation by acquiring 1-bit measurements and train a long-short term memory (LSTM) NN to improve the reconstruction of the channel phase. 
The main contributions of this paper are the combination of 
cGAN and LSTM for stable channel estimation in massive MIMO scenarios with mixed resolution ADCs. Moreover, we adopt Unitary tensor-ESPRIT~\cite{08HaardtTensor} for super resolution parameter estimation and use this as a baseline for our results. 

In this paper, 
Section~\ref{sec:channel} presents our system overview and the wireless channel model, Section~\ref{sec:method} introduces our proposed method, Section~\ref{sec:cGAN} presents details about our cGAN, Section~\ref{sec:lstm} shows the processing performed at the LSTM, 
Section~\ref{sec:results} presents our results, and Section~\ref{sec:conclusion} concludes our paper. 

\begin{figure}[tb!]
    \centering
    \includegraphics[width=0.85\columnwidth]{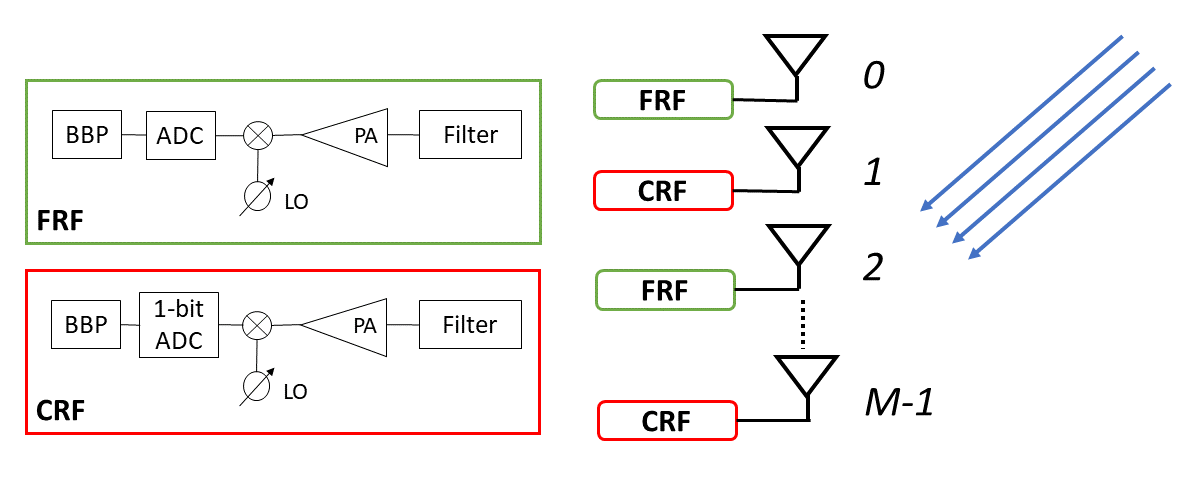}
    \caption{Overview of the configuration of RF chains at BS for our tackled problem. Even/odd antenna elements are connected to FRFs/CRFs, respectively.}
    \label{fig:systemview}
\end{figure}

\section{System and Channel Model}
\label{sec:channel}

Figure~\ref{fig:systemview} presents a block diagram 
of the full resolution RF chains (FRFs) and constrained RF chains (CRFs) with 
which the BS is equipped. Analog filters, power amplifiers, local oscillators, and baseband processing units are available in both types of RF chains. However, they differ with respect to their ADC resolution. We apply ML methods to tackle the problem of uplink channel estimation for a mixed resolution receiver. Hence,
we define three channel matrices: $\mathbf{H}$ is our desired channel, $\mathbf{H}_\mathrm{ce}$ is the input to our first ML instance, and $\mathbf{H}_{c}$ is the input to Unitary tensor-ESPRIT. 

First, we consider the wireless channel $\mathbf{H} \in \mathbb{C}^{M \times N_\mathrm{sub}}$ 
of an orthogonal frequency division multiplexing (OFDM) system
in the spatial and frequency domain with $N_\mathrm{sub}$ sub-carriers and $M$ antenna elements at the BS. 

The channel in each sub-carrier $\mathbf{h}(n)$ is modeled as      
\begin{equation}
\mathbf{h}(n) = \sum_{i=1}^{L} \alpha_i e^{-j 2 \pi \frac{(n-1)}{N_\mathrm{sub}} \tau_i} \mathbf{a}_F(\theta_i,d,M), 
\label{eq:channel}
\end{equation} 
where $L$ is the number of multipath components (MPCs), $\tau_i, \alpha_i$, and $\theta_i$ are, respectively, the delay, complex amplitude, and direction of arrival (DoA) of each $i^\mathrm{th}$ MPC. The uniform linear array (ULA) steering vector $\textbf{a}_F$ at the BS is modeled as
\begin{equation}
\textbf{a}_F(\theta_i,d,M)= [1, e^{j\mu_i}, e^{j2\mu_i}, \ldots e^{j(N_\mathrm{ant}-1)\mu_i}]^T,   
\label{eq:steer}
\end{equation} where 
$ \mu_i = \frac{2\pi}{\lambda}d \cos{\theta_i}$, is the spatial frequency and $d=\frac{\lambda}{2}$ is the spacing between the antenna elements. 

Equation~(\ref{eq:steer}) models a full resolution array steering vector in which each antenna element is connected to one RF chain, and we use full resolution ADCs.  
For reduced resolution arrays, we assume that not all antenna elements are connected to a full resolution RF chain.
Without loss of generality, we assume that every odd 
antenna element is connected to a low resolution RF chain. Our proposed method first 
uses only the FRF and
ignores the information from the CRF.
Therefore, we assume $M'= \frac{M}{2}$ and $d'=\lambda$ 
to derive the constrained wireless channel response $\mathbf{h}_c(n)$ per sub-carrier 
\begin{equation}
\mathbf{h}_c(n) = \sum_{i=1}^{L} \alpha_i e^{-j 2 \pi \frac{(n-1)}{N_\mathrm{sub}} \tau_i} \mathbf{a}_F(\theta_i,d',M') + \mathbf{z}(n),
\label{eq:conschannel}
\end{equation} 
where $\mathbf{z}(n) \in \mathbb{C}^{N_\mathrm{ant}'}$ is zero mean circular symmetric Gaussian noise, and $\mathbf{H}_c = [\mathbf{h}_c(0), \mathbf{h}_c(1), \ldots, \mathbf{h}_c(N_\mathrm{sub}-1)] \in \mathbb{C}^{M' \times N_\mathrm{sub}}$. 
In order to make the dimensionality of $\mathbf{H}_c$ equal to the dimensionality of $\mathbf{H}$,  
$\mathbf{H}_c$ is expanded to $\mathbf{H}_\mathrm{ce}$ by inserting zero row vectors $\mathbf{0}$ in the odd row antenna positions, then $\mathbf{H}_\mathrm{ce} \in \mathbb{C}^{M \times N_\mathrm{sub}}$.

\section{Two-step ML approach for channel estimation} 
\label{sec:method}

\begin{figure}[tb!]
    \centering
    \includegraphics[width=0.9\columnwidth]{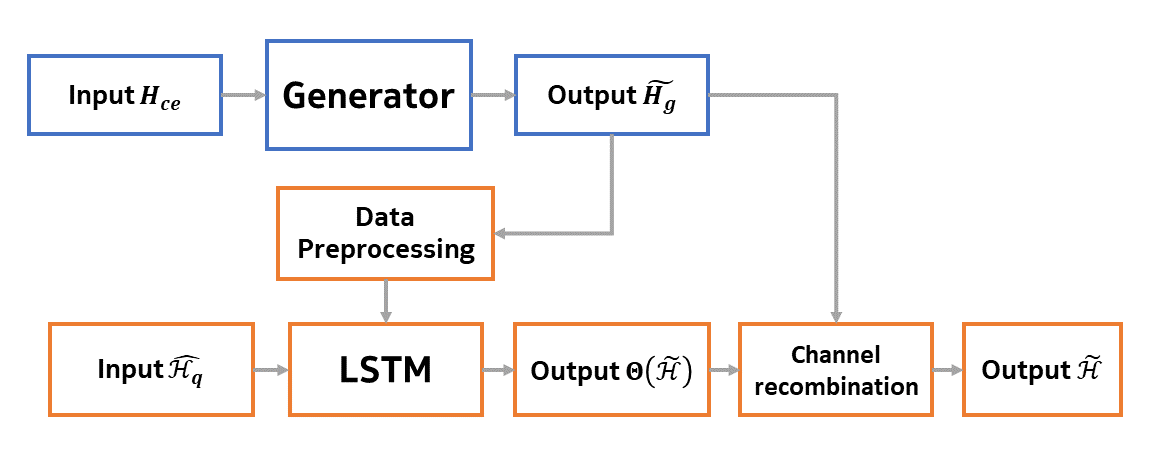}
    \caption{Two-step ML approach for channel estimation in massive MIMO scenario with antenna elements connected to mixed resolution RF chains. Blue and orange highlight the first and second ML steps, respectively.}
    \label{fig:general-block}
\end{figure}

Figure~\ref{fig:general-block} summarizes our proposed method for channel estimation for massive MIMO systems, where the BS is equipped with full resolution RF chains for every even and constrained ones for every odd antenna.   
The first and second ML instances are highlighted in blue and orange, respectively. Our first ML instance relies only on measurements from the antenna elements connected to full resolution RF chains. Here, we train a cGAN
to predict the channel of the antenna elements with low resolution RF chains and estimate the channel of the antenna elements with full resolution RF chains. In our second ML instance, we consider the low resolution measurements and 
the results from the 
cGAN as input to a LSTM NN in order to improve the phase accuracy of the channel estimation. After the LSTM, its phase output is combined with the preprocessed absolute value of the cGAN's output. This is our final complex valued channel estimate $\mathbf{\tilde{H}}$ in the time domain. 
Since 1-bit (for real and imaginary parts separately) quantized signals are similar to quadrature phase shift keying (QPSK) symbols, only phase information can be extracted from such measurements. Therefore, our second ML instance only aim to improve the phase signal.
The following sections explain in detail how each ML instance operates.  

\section{Channel estimation with cGAN}
\label{sec:cGAN}
\begin{figure}[bt!]
\centering
\includegraphics[width=0.9\columnwidth]{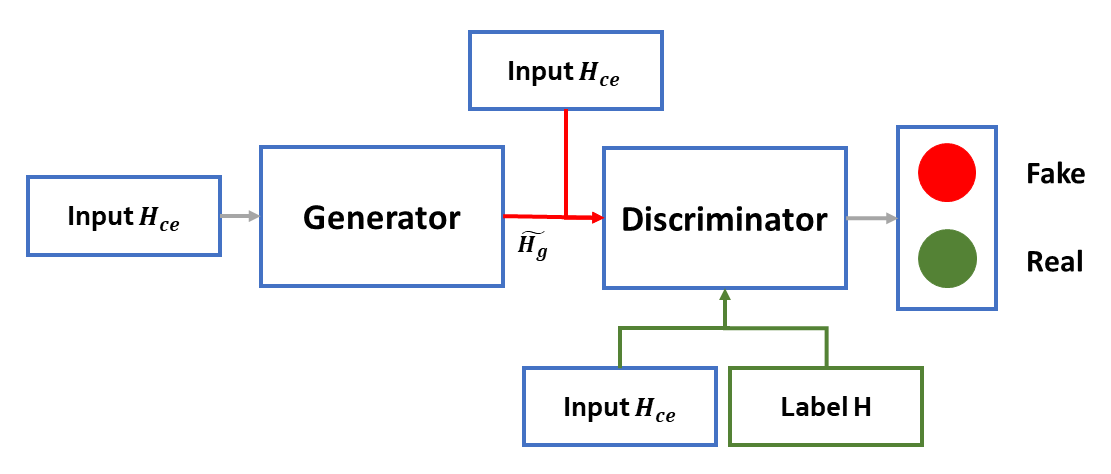}
\caption{Conditional GAN, two NNs play a minmax game where the generator tries to fool the discriminator. Both NNs have knowledge of the prior information $\mathbf{H}_\mathrm{ce}$. The discriminator should classify  $[\mathbf{\tilde{H}}_g,\mathbf{H}_\mathrm{ce}]$ as a fake sample, while $[\mathbf{H},\mathbf{H}_\mathrm{ce}]$ is classified as a real sample. The generator fools the discriminator when $[\mathbf{\tilde{H}}_g,\mathbf{H}_\mathrm{ce}]$ is classified as real~\cite{14GoodfellowGAN}.}
\label{fig:cgan}
\end{figure}

Inspired by the success of 
cGANs on image to image translation,
we employ a similar architecture as the \textit{Pix2Pix} application ~\cite{17IsolaPix}  
for tackling the problem of channel estimation with mixed resolution RF chains.
Figure~\ref{fig:cgan} depicts the interplay between the two NNs in the cGAN training phase.  
This section starts by presenting our dataset preprocessing. In the following, we discuss our cGAN architecture, comment on the adversarial training, and its optimization function. 

\subsection{Dataset preprocesing for cGAN}
\label{sub:datagan}
For the first ML instance 
of our 
combined approach, 
$\mathbf{H}_\mathrm{ce}$ is used as input and $\mathbf{H}$ is the desired output or label. However, DL 
libraries do not work with complex values. Moreover,  
the input/output coefficients should be limited to a known range of values to improve convergence. Therefore, a preprocessing is employed as
\begin{itemize}
    \item $\mathbf{H}$ is normalized by its Frobenius norm, and then multiplied by a scaling factor 
    to increase the range value of the channel coefficients without changing their statistical distribution.
    \item $\mathbf{H} \in \mathbb{C}^{M \times N_\mathrm{sub}}$ is rearranged by concatenating $\mathfrak{Re}\{\mathbf{H}\}$ and $\mathfrak{Im}\{\mathbf{H}\}$ in their third dimension. 
     
\end{itemize}
The same preprocessing is performed for $\mathbf{H}_\mathrm{ce}$  
Therefore, the input $\mathbf{H}_\mathrm{ce}$ and label $\mathbf{H}$ of our first ML 
problem are 3-dimensional with size $[M \times N_\mathrm{sub} \times 2]$. 

\subsection{Adversarial Network Architecture}
\begin{figure}[tb!]
\centering
\includegraphics[width=\columnwidth]{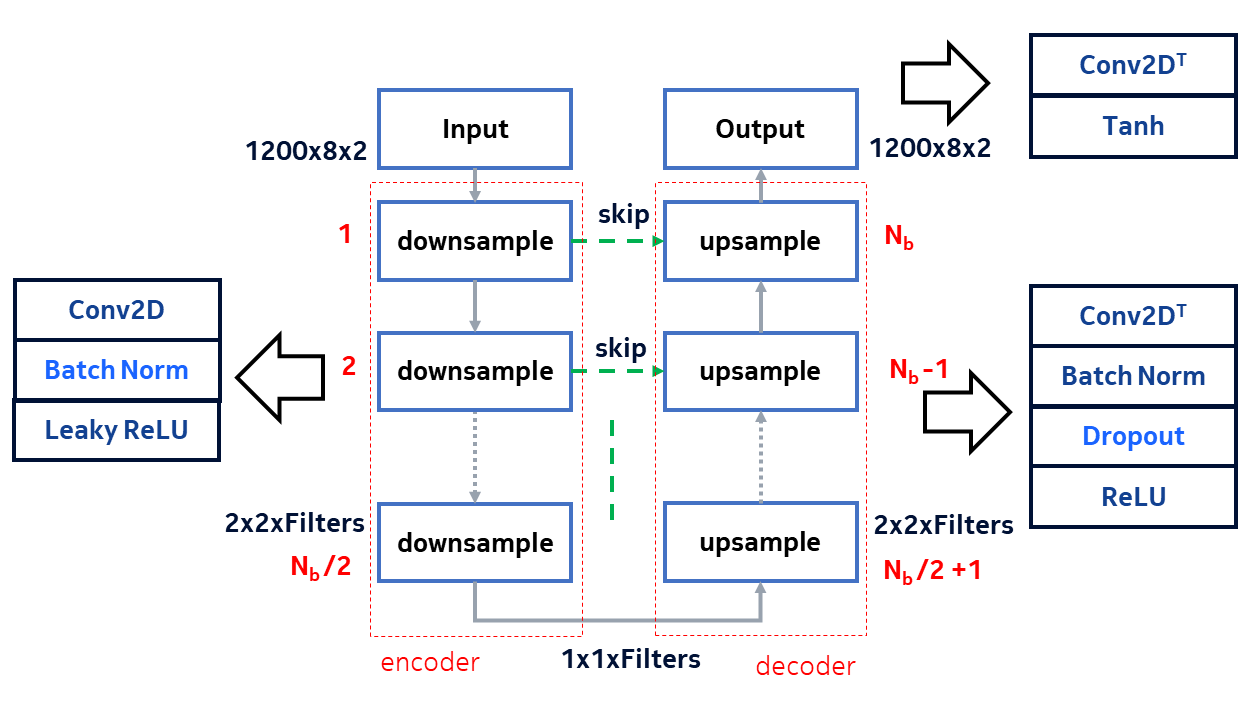}
\caption{U-Net architecture deployed as the generator including encoder and decoder pipeline and numbering for skip connections.}
\label{fig:u-net}
\end{figure}

As shown in Figure~\ref{fig:cgan}, two NNs are deployed for the adversarial training. Here, 
the generator NN consists of a U-shaped deep NN which works similar to an encoder-decoder architecture, but includes skip connections between blocks $j$ and $N_b-j$, where $j=[1:N_b]$, and $N_b$ is the total number of processing blocks, see Figure~\ref{fig:u-net}. Those block interconnections provide more information to the decoder block which receives data from the encoder-decoder pipeline and its related layer in the encoder side~\cite{17IsolaPix}. For our tackled problem, the skip connections helps to recover the FRF estimation since they are the underlying data structure from the input that we want to keep in the output of the generator NN.

\begin{figure}[tb!]
\centering
\includegraphics[width=0.9\columnwidth]{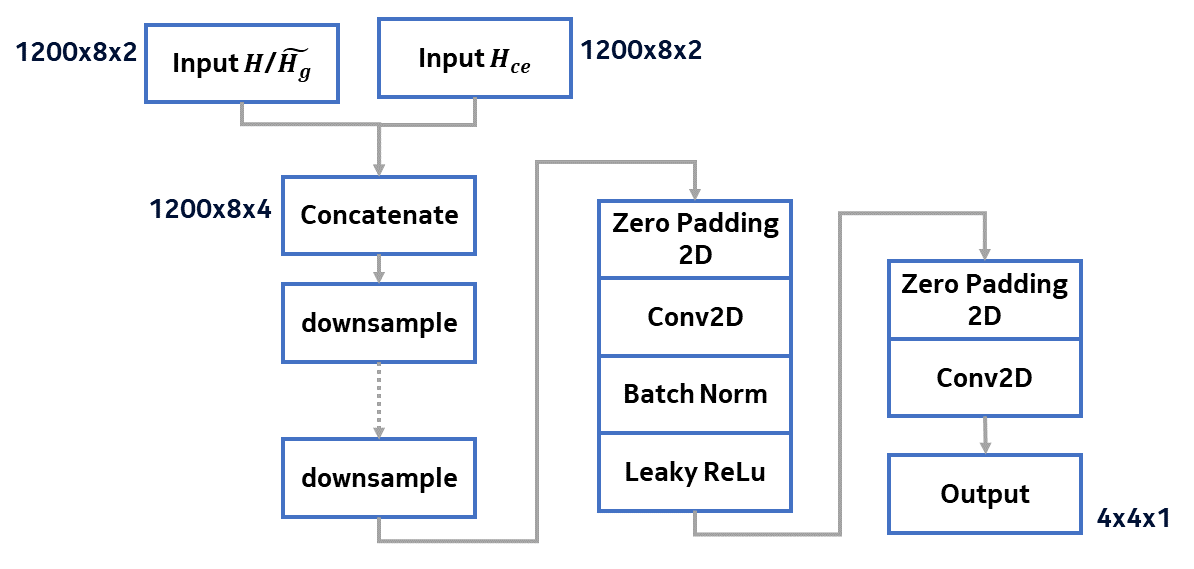}
\caption{Patch-Net architecture deployed for the discriminator.}
\label{fig:patch-net}
\end{figure}

Figure~\ref{fig:u-net} shows the U-Net architecture employed for the generator, the downsample blocks from  the encoder and the upsample blocks from the decoder. Each downsample block consists of one convolutional 2-dimensional layer (Conv2D), one batch normalization layer (BatchNorm), and a leaky rectifier linear unit (LeakyReLU) activation function, where $y = x$ for $x > 0$, and $y = 0.3x$ for $x<0$. Each upsample block consists of one transposed convolutional 2-dimensional layer (Conv2D$^T$), followed by BatchNorm and rectifier linear unit (ReLU) as activation function. The architecture structure was inspired by~\cite{17IsolaPix}; however, here we adapt the filter length and how it is shifted (stride) in the convolutional layers. This is done to reduce the input size to $[1 \times 1 \times N_\mathrm{filter}]$ after the $N_b/2$ downsampling processing blocks, where $N_\mathrm{filter}$ is the number of filters in the previous convolutional layer.

Figure~\ref{fig:patch-net} presents the discriminator NN, called Patch-NN, which reduces the size of the input to $N \times N$, where $N$ is the size of the patch, and classifies each coefficient as real or fake.  
For that, first the discriminator concatenates the conditional input $\mathbf{H}_\mathrm{ce}$ to the label $\mathbf{H}$ or to the generated channel $\mathbf{\tilde{H}}_g$, forming, respectively, the 
real and fake classes. Then, the input is downsampled by the 
downsampling blocks which are followed by one zero padding layer, and one Conv2D with BatchNorm and LeakyReLU activation function. After one zero padding and one Conv2D layer, the discriminator provides its output of size $[N \times N \times 1]$. During
the optimization, this output is further averaged and represented as a scalar value~\cite{17LedigPhoto}.

\subsection{Optimization with cGAN}
In cGAN, two NNs play a minmax game where the generator tries to fool the discriminator, and it is conditional because some prior knowledge is provided. Mathematically, the loss function of a cGAN is 
\begin{equation}
\begin{aligned}
\mathcal{L}_{\mathrm{cGAN}}(G,D) = & \mathbb{E}_{x,y}[\log D(x,y)] + \\
& \mathbb{E}_{x,z}[\log (1-D(x,G(x,z)))],
    \label{eq:cgan-loss}
\end{aligned}
\end{equation}
where the generator $G$ learns to map input data $x$ and random noise $z$ to output data $y$, $G:\{ x,z\} \rightarrow  y$, and the discriminator $D$ tries to recognize 
the channels generated by $G$. In order to have the
generated
output wireless channels $\mathbf{\tilde{H}}_g$ close to the wireless channel labels $\mathbf{H}$, a weighted $L_2$ term 
\begin{equation}
\mathcal{L}_{\mathrm{L_2}}(G) =  \mathbb{E}_{x,y,z}[\| y-G(x,z) \|_2]
    \label{eq:generator-l1}
\end{equation}
is included on the generator loss function. Therefore, the final optimization objective is 
\begin{equation}
G^* = \arg \min_{G} \max_{D}  \mathcal{L}_{\mathrm{cGAN}}(G,D) + \beta \mathcal{L}_{\mathrm{L_2}},
    \label{eq:final-loss}
\end{equation}
where $\beta$ is the weighting factor. 

The generator and discriminator NNs are trained together in each epoch. For testing, or inference, only the generator architecture is used. Therefore, only knowledge of $\mathbf{H}_\mathrm{ce}$ is needed. 

\section{Phase improvement with LSTM}
\label{sec:lstm}

In this section we enhance the
channel estimation based on quantized measurements from constrained RF-chains. 
For that purpose, we
develop a second ML step based on LSTM to improve the phase of the channel estimated by our cGAN. This section presents the data preprocessing to obtain the inputs to the LSTM NN, the equations describing its architecture and comment on the cost function.   

\subsection{Data preprocessing for LSTM}
\label{sub:datalstm}
After estimating $\mathbf{\tilde{H}_g}$ with
our first ML instance, we consider 
the quantized measurements of the antenna elements with low resolution RF chains, 1-bit ADCs in this case. As the 1-bit RF chains are of low cost and size, we assume each of the $M$ antennas
can perform 1-bit measurements. 
As our second ML instance operates in the time domain, the following data manipulation are made: computation of the inverse Fourier transform $\mathscr{F}^{-1}$, profiling computation~\cite{20ZirwasProfiling}, random sequence mixture, and 1-bit quantization.

First, the full resolution channel impulse responses (CIRs) 
$\mathcal{H}_x~\in~\mathbb{C}^{M \times N_\mathrm{sub}}$  are computed for the desired, noisy measurement and predicted signals, respectively, as $\mathbf{\mathcal{H}} = \mathscr{F}^{-1}(\mathbf{H}), ~\mathbf{\mathcal{H}}_z = \mathscr{F}^{-1}(\mathbf{H} + \mathbf{Z}),~\mathrm{and}~ \mathbf{\mathcal{H}}_g = \mathscr{F}^{-1}(\mathbf{\tilde{H}}_g)$, being $\mathbf{Z}$ the additive white Gaussian noise. 
Second, we compute the profiling version of each full resolution signal. The profiling operation, best described in~\cite{20ZirwasProfiling}, consists, basically, in oversampling the CIR. At this point, we also reduce the signal observation window to $K<N_\mathrm{sub}$, so that $\mathcal{H}_x~\in~\mathbb{C}^{M \times K}$. This windowing is mainly performed due to the small number of MPCs and their clustering, which leads the CIR to contain relevant power in a limited number of taps $K$. Hence, this filtering also limits the impact of noise.

For those 1-bit measurements, we assume the 
user equipment (UE) sends the pilots mixed with a random sequence 
$\mathbf{S} \in \mathbb{C}^{K \times 1}$, generated in the frequency domain with constant amplitude and random phase. The random sequence $\mathbf{S}$ is used to spread the amplitude information within the measurement window which increases the useful information contained in the phase signal. 
Third, the profiled signals are transferred to the frequency domain by the Fourier transform $\mathscr{F}$, and multiplied with the random sequence $\mathbf{S}$; then, they are transformed back to the time domain by an $\mathscr{F}^{-1}$. This set of operations is summarized as

\begin{equation}
    \hat{\mathcal{H}}_x = \mathscr{F}^{-1} \{ \mathscr{F}\{P(\mathcal{H}_x)\} \mathbf{S} \},
    \label{eq:profileseq}
\end{equation}
where $x$ refers, generically, to our signals of interest, and $P$ is the profiling operation. Finally, the 1-bit measurements used in the second ML instance,
are composed as
\begin{equation}
    \hat{\mathcal{H}}_q = Q(\mathfrak{Re}\{ \hat{\mathcal{H}}_z\}) + j~ Q(\mathfrak{Im}\{ \hat{\mathcal{H}}_z\}),
\end{equation}
where $Q$ is the quantization function.  

The objective of the second ML instance
is to improve the phase signal estimation in the full resolution time domain. Therefore, the input is a combination of full-resolution $\hat{\mathcal{H}}_g$ and 1-bit $\hat{\mathcal{H}}_q$ signals. Specifically, the inputs are  $\mathcal{I}_1 = \Theta(\hat{\mathcal{H}}_g)$, and  $\mathcal{I}_2 = \Theta(\hat{\mathcal{H}}_q)$,
where $\Theta$ is the angle operation and $\mathcal{I}_x \in \mathbb{R}^{K \times 1}$. As every antenna element phase signal is optimized separately, the dataset size is increased by $M$. $\Theta(\hat{\mathcal{H}})$ is the label for our LSTM training.   

\subsection{LSTM architecture and cost function}

\begin{figure}[tb!]
  \centering
  \includegraphics[width=0.7\columnwidth]{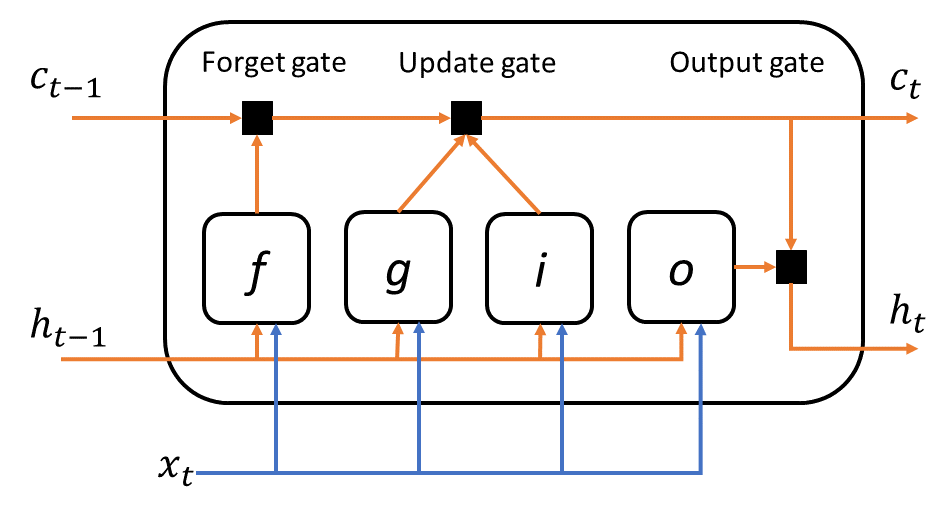}
   \caption{LSTM unit, set of computations performed for each time step of an input sequence $x_t$.}
    \label{fig:lstm_cell}
\end{figure}

LSTM NNs are mainly used when the data 
has some time dependency. For instance, speech recognition, natural language processing, and time forecast, to name a few. A LSTM maps an input sequence $x = (x_1, x_2, \ldots, x_K)$ to an output sequence $y = (y_1, y_2, \ldots, y_K)$ by computing iteratively ($t=1:K$) the following equations ~\cite{97HochreiterLSTM}, 
\begin{align}
    i_t = \sigma(\mathbf{W}_{i}x_t + \mathbf{R}_i h_{t-1} + \mathbf{b}_i) \\
    f_t = \sigma(\mathbf{W}_{f}x_t + \mathbf{R}_f h_{t-1} + \mathbf{b}_f) \\
    g_t = \sigma(\mathbf{W}_{g}x_t + \mathbf{R}_g h_{t-1} + \mathbf{b}_g) \\
    o_t = \sigma(\mathbf{W}_{o}x_t + \mathbf{R}_o h_{t-1} + \mathbf{b}_o)\\
    c_t = f_t \odot c_{t-1} + i_t \odot g_t\\
    h_t = o_t \odot \theta(c_t)\\
    y = \phi(\mathbf{W}_y h_t + \mathbf{b}_y),
\end{align}
where $\mathbf{W}_x, \mathbf{R}_x$ and $\mathbf{b}_x$ refer, generically, to input weight matrices, recurrent weight matrices, and biases vectors, respectively. The LSTM processing parts are: $i$ the input gate, $f$ the forget gate, $g$ the cell candidate, $o$ the output gate,  
$c$ the cell state, $h$ the hidden state, and $y$ the decided output sequence. The parameters $\sigma, \theta$ and $\phi$ are the activation functions, while $\odot$ is the element-wise product.
Figure~\ref{fig:lstm_cell} shows a block diagram of a single LSTM unit.

The LSTM for channel phase improvement is trained in a supervised learning setting, using the mean square error (MSE) as the cost function which should be minimized by the 
truncated back-propagation through time (BPTT) algorithm~\cite{97HochreiterLSTM}.

\section{Simulation and Results}
\label{sec:results}

\begin{table}[bt!]
\centering
\caption{Description of the U-Net deployed as generator NN.}
\label{tab:gen-desc}
\resizebox{0.7\linewidth}{!}{%
\begin{tabular}{|c|c|c|c|c|c|}
\hline
$j$ & Block & $N_\mathrm{filter}$ & Stride & BatchNorm & Dropout \\ \hline
1 & downsample & 32 & [1,1] & No & No \\ \hline
2 & downsample & 64 & [2,2] & Yes & No \\ \hline
3 & downsample & 64 & [2,2] & Yes & No \\ \hline 
4 & downsample & 64 & [2,2] & Yes & No \\ \hline 
5 & downsample & 64 & [5,1] & Yes & No \\ \hline
6 & downsample & 64 & [5,1] & Yes & No \\ \hline
7 & downsample & 128 & [6,1] & Yes & No \\ \hline 
8 & upsample & 128 & [6,1] & Yes & Yes \\ \hline 
9 & upsample & 64 & [5,1] & Yes & Yes \\ \hline
10 & upsample & 64 & [5,1] & Yes & Yes \\ \hline
11 & upsample & 65 & [2,2] & Yes & No \\ \hline
12 & upsample & 128 & [2,2] & Yes & No \\ \hline
13 & upsample & 64 & [2,2] & Yes & No \\ \hline
14 & upsample & 32 & [1,1] & Yes & No \\ \hline
\end{tabular}}
\end{table}

\begin{table}[bt!]
\centering
\caption{Description of the Patch-Net deployed as discriminator NN.}
\label{tab:dis-desc}
\resizebox{0.7\linewidth}{!}{%
\begin{tabular}{|c|c|c|c|c|}
\hline
 Block & $N_\mathrm{filter}$ & Stride & BatchNorm & Activation \\ \hline
 downsample & 64 & [1,1] & No & - \\ \hline
 downsample & 128 & [5,1] & Yes & - \\ \hline
 downsample & 128 & [5,1] & Yes & - \\ \hline
 downsample & 128 & [3,1] & Yes & - \\ \hline
 downsample & 128 & [2,1] & Yes & - \\ \hline
 zero padding 2D & - & - & - & -  \\ \hline
 Conv2D & 256 & [1,1] & Yes & LeakyReLU \\ \hline
 zero padding 2D & - & - & - & - \\ \hline
 Conv2D & 1 & [1,1] & No & Linear \\ \hline
 \end{tabular}}
\end{table}

\begin{table}[tb!]
\centering
\caption{Description of the LSTM NN.}
\label{tab:lstm-struc}
\resizebox{0.7\linewidth}{!}{%
\begin{tabular}{|c|c|c|c|}
\hline
 Layer & $N_\mathrm{filter}$ & Return Sequences & Activation \\ \hline
 LSTM & 10 & Yes & hyperbolic tangent \\ \hline
 LSTM & 1 & Yes & linear \\ \hline 
 \end{tabular}}
\end{table}

First, we model two different datasets, dataset 1 has channels with 3 MPCs and dataset 2 has wireless channels with 5 MPCs, both modeled by 
Equation~(\ref{eq:channel}). The channel datasets follow a Rayleigh distribution, and the delays and 
DoAs are drawn from a uniform distribution. All the MPCs within a dataset have different delay values, and
a maximum delay of about 163~ns. Each dataset is of size 1500, and the desired wireless channels are parameterized by $M=8,~d=\lambda/2$, $N_\mathrm{sub}=1200$. For the measured 
signals, we consider a SNR of $20$~dB.

Unitary tensor-ESPRIT~\cite{08HaardtTensor} is selected as baseline for performance comparison. 
The input $\mathbf{H}_c$ to the Unitary tensor ESPRIT algorithm is parameterized by
$M'= 4,~d'=\lambda$. 
Aiming to keep the $1:1$~relationship between spatial frequencies and DoAs, we limit the DoA $\in [0:\pi/4]$. 
The results for Unitary tensor-ESPRIT are computed separately for each dataset, as it needs to know the correct number of MPCs in advance to compute the channel parameters (DoAs, delays, and complex amplitudes).
After the channel parameter estimation, the wireless channel is recovered in the frequency domain and undergoes similar processing as in equation~(\ref{eq:profileseq}). 
The results are compared by means of the cumulative distribution function (CDF) of the 
normalized squared error (NSE) which is computed, generically, as 
    $\mathrm{NSE} = \frac{\|\mathbf{A} -\mathbf{\tilde{A}}  \|_F^2}{\|\mathbf{A}\|_F^2}$ for matrices.

\begin{figure}[tb!]
\centering
\includegraphics[width=\columnwidth]{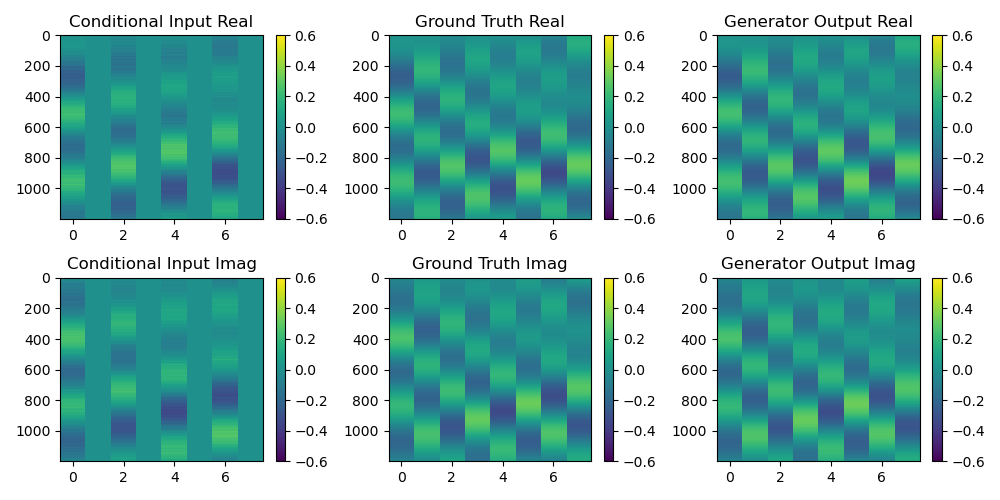}
\caption{Channel estimation using cGAN, sample result where x-axis is the number of antennas, y-axis is the number of sub-carriers. The matrices in the first column are the input $\mathbf{H}_\mathrm{ce}$ with odd antenna elements zeroed, the second column show the ground truth wireless channel $\mathbf{H}$, and the third column shows the estimated channel $\mathbf{\tilde{H}}_g$ after 150 epochs of training.}
\label{fig:sample-result}
\end{figure}

Regarding our proposed 2-step machine learning approach, first we train the cGAN and after that the LSTM. 
Both trainings are performed using TensorFlow 2.0, Keras and Pyhton. For cGAN, we start by preprocessing the dataset as described in 
Subsection~\ref{sub:datagan}.
The input to our generator is $\mathbf{H}_\mathrm{ce}$. The generator consists of 7 downsampled blocks and 7 upsampled blocks with a skip connection between blocks $j$ and $N_b-j$, where 
$N_b = 14$. 
The filter length for all convolution operations is set to 5. Details on $N_\mathrm{filter}$, stride size and dropout usage are provided in Table~\ref{tab:gen-desc}. 
The input to our discriminator are $[\mathbf{H}_\mathrm{ce}, \mathbf{H}]$ for true case, and $[\mathbf{H}_\mathrm{ce}, \tilde{\mathbf{H}}_g]$ for fake case. The discriminator reduces the input channel dimensions to $4 \times 4$ by applying 5 downsampling blocks; 
a detailed description of our discriminator architecture is presented in Table~\ref{tab:dis-desc}.
The layers weights, for both generator and discriminator, are initialized from a normal distribution with zero mean and $\sigma = 0.2$ standard deviation. The Adam optimizer~\cite{14KingmaAdam} with $2 \times 10^{-4}$ initial learning rate is used for both NNs, and $\beta=100$. The adversarial training runs for 150 epochs, with only 600 dataset samples, $50\%$ from dataset 1, and $50\%$ from dataset 2. The remaining 2400 dataset samples are used for testing. 
Figure~\ref{fig:sample-result} presents a test sample result for our channel estimation approach employing our cGAN. 

After the channel estimation by the cGAN,  
the second ML instance runs 
to improve the estimated phase signal. For this purpose, the dataset is preprocessed as in Subsection~\ref{sub:datalstm}, and a two layers LSTM NN is deployed as described in Table~\ref{tab:lstm-struc}. The learning rate is set to $2 \times 10^{-4}$, and the gradient descent optimizer is Adam~\cite{14KingmaAdam}. Then, the supervised training runs for 50 epochs, using a training dataset with 2400 samples, where $50\%$ are from dataset 1, and $50\%$ are from dataset 2. The remaining 600 combined dataset samples are used for testing.  
 
\begin{figure}[bt!]
    \centering
    \includegraphics[width=0.9\columnwidth]{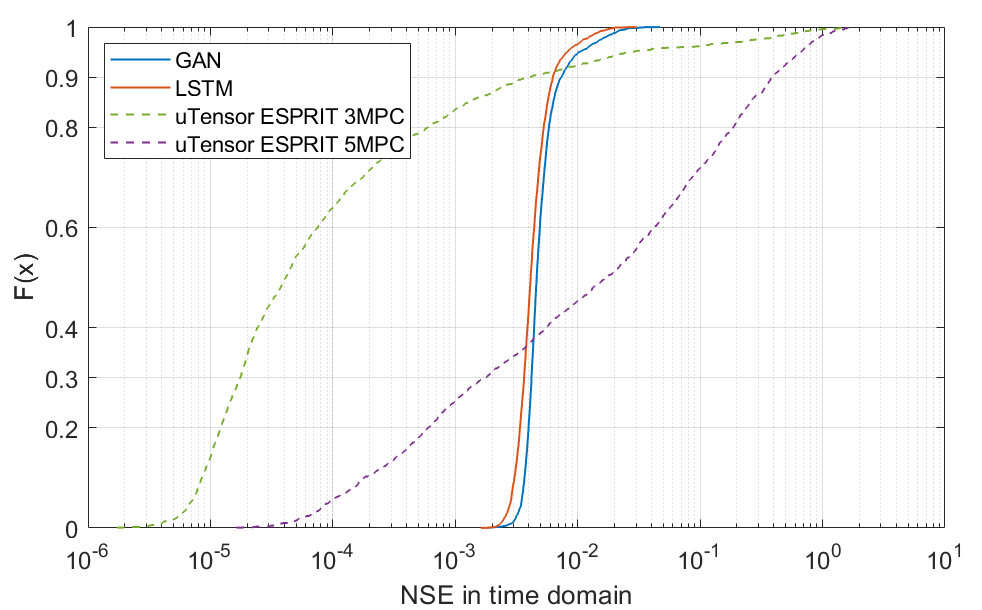}
    \caption{Comparison of the NSE in the time domain for cGAN, LSTM and Unitary tensor-ESPRIT. All channel estimations considered SNR$~=~20$dB.}
    \label{fig:timeNSE}
\end{figure}

Figure~\ref{fig:timeNSE} presents the comparison of Unitary tensor-ESPRIT, our cGAN, and LSTM. 
For Unitary tensor-ESPRIT at 3~MPC and 5~MPC, there is a large variation on the NSE. This is due to the delay spacing between the MPCs. For the 3~MPCs dataset, Unitary tensor-ESPRIT has better channel estimation error in slightly more than $90\%$ of the channels.  
However, for the 5~MPCs dataset, less than $40\%$ of the channels are better estimated by Unitary tensor-ESPRIT. 
Since the cGAN does not rely on parameter estimation, but, instead, tries to model the dataset statistical distribution, it is  successful in reconstructing channels with various, closely spaced, MPCs. Moreover,  
the cGAN gives a stable reconstruction error. 
Such small variation on the cGAN NSE performance  
indicates that our generator architecture was capable to generalize and map the dataset distribution. It is also important to point out that the cGAN estimation error is below the measurement error in around $95\%$ of the cases.
Building upon the good results of our cGAN,  
the LSTM has improved the channel estimates by $0.6$~dB in average, with only 568 trainable parameters. 
This means that quantized measurements also add useful information to ML problems. 

\section{Conclusion}
\label{sec:conclusion}
In this paper we propose 
a two-step ML 
approach for channel estimation 
in massive MIMO systems
with mixed resolution RF chains. 
The results show that this combined approach is independent of the number of MPCs and stable. Mainly, it is competitive with Unitary tensor-ESPRIT  
when the channel has many closely spaced MPCs.  
Future work may consider a 
joint operation between parameter estimation algorithms and our two-step ML 
method. 

\section*{Acknowledgement}
This research was partly funded by German Ministry of Education and Research (BMBF) under grant 16KIS1184 (FunKI).

\bibliographystyle{IEEEtran}
\bibliography{IEEEabrv,mybibfile}

\end{document}